\documentclass[aps,prd,preprintnumbers,twocolumn,superscriptaddress,nofootinbib]{revtex4-1}
\usepackage{lmodern}
\usepackage{hyperref}
\hypersetup{colorlinks=true,linkcolor=purple,anchorcolor=blue,citecolor=blue, filecolor=blue,urlcolor=blue,bookmarksnumbered=true,
pdfview=FitB
}
\usepackage{mathrsfs}
\usepackage{rsfso}
\usepackage{color}
\usepackage{xcolor}
\usepackage{ulem}
\usepackage{csquotes}
\colorlet{purple1}{blue!70!red}
\colorlet{darkred}{red!50!black}

\usepackage{graphicx}
\usepackage[bf,SL,BF]{subfigure}
\usepackage{psfrag}
\usepackage{color}
\usepackage{amssymb}
\usepackage{amsmath}
\usepackage{epstopdf}
\usepackage{natbib}
\usepackage{amssymb}
\usepackage{amsmath,amssymb}
\usepackage{mathtools}
\usepackage{float}
\usepackage{color}
\usepackage{leftidx}
\usepackage{booktabs}
\usepackage{latexsym}
\usepackage{revsymb}
\usepackage{multirow}
\usepackage{hypcap}
\usepackage{array}
\usepackage[english]{babel}
\usepackage{amsmath}
\usepackage{cleveref}
\def\orcid#1{\kern .08em\href{https://orcid.org/#1}{\includegraphics[keepaspectratio,width=0.7em]{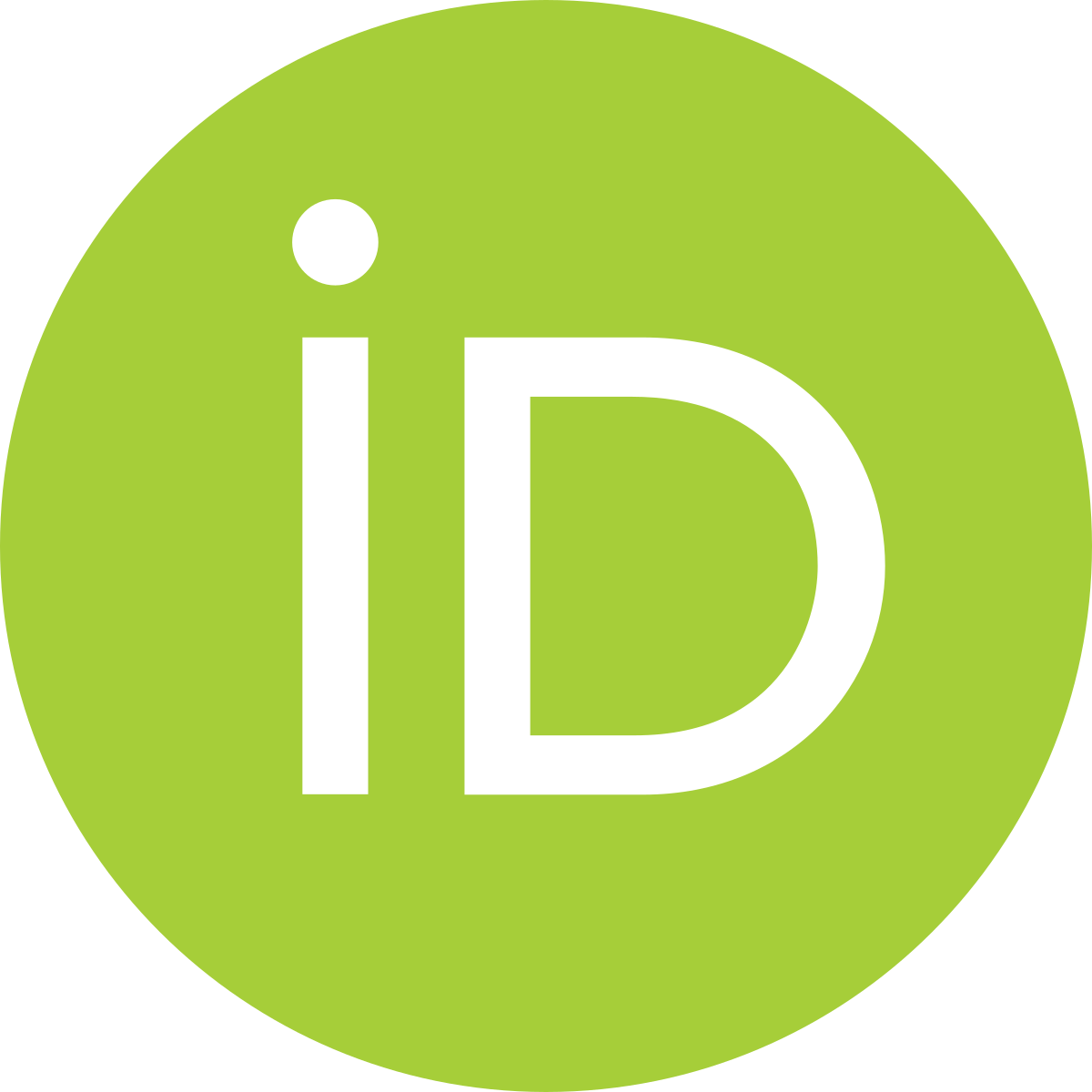}}}
\newcommand{\be}{\begin{eqnarray}}
	\newcommand{\ee}{\end{eqnarray}}
	
\def\orcid#1{\kern .08em\href{https://orcid.org/#1}{\includegraphics[keepaspectratio,width=0.7em]{ORCID_iD.png}}}

\newcommand{\bfq}{{\bf q}_{\perp}}

\newcommand{\bfk}{{\bf k}_{\perp}}

\newcommand{\kp}{{k}_{\perp}}
\newcommand{\bfb}{{\bf b}_{\perp}} 
 
\newcommand{\bp}{{b}_{\perp}}

\begin{document}

	\title{Azimuthal spin asymmetries in pion-polarized proton induced Drell-Yan process at COMPASS using holographic light-front QCD}	
	
	\author{Bheemsehan~Gurjar\orcid{0000-0001-7388-3455}}
	\email{gbheem@iitk.ac.in} 
	\affiliation{Indian Institute of Technology Kanpur, Kanpur-208016, India}
	
	\author{Chandan~Mondal\orcid{0000-0002-0000-5317}}
		\email{mondal@impcas.ac.cn} 
	\affiliation{Institute of Modern Physics, Chinese Academy of Sciences, Lanzhou 730000, China}
	\affiliation{School of Nuclear Science and Technology, University of Chinese Academy of Sciences, Beijing 100049, China}

	\date{\today}
\begin{abstract}
	We compute all the leading-twist azimuthal spin asymmetries in the pion-proton induced Drell-Yan process. These spin asymmetries arise from convolutions of the leading-twist transverse-momentum-dependent parton distributions (TMDs) of both the incoming pion and the target proton. We employ the holographic light-front pion wave functions for the pion TMDs, while for the proton TMDs, we utilize a light-front quark-diquark model constructed by the soft-wall AdS/QCD. The gluon rescattering is crucial to predict nonzero time-reversal odd TMDs. We study the utility of a nonperturbative SU$(3)$ gluon rescattering kernel, which extends beyond the typical assumption of perturbative U$(1)$ gluons. Subsequently, we employ Collins-Soper scale evolution at Next-to-Leading Logarithmic precision for the TMDs evolution. Our predictions for the spin asymmetries are consistent with the available experimental data from COMPASS and other phenomenological studies. We also present the cross section for the pion-nucleus induced Drell-Yan process with the obtained pion TMDs supplemented by the TMDs of the target nucleus.
\end{abstract}
\maketitle
%
\section{Introduction}\label{sec:Intro}
Over the last two decades, studying dilepton production in high-energy hadron-hadron collisions, known as the Drell-Yan (DY) process~\cite{Drell:1970wh,drell2000partons}, has emerged as a crucial method for gaining insights into the fundamental structure of hadrons, notably through the analysis of parton distributions.
The differential cross-section of the DY process follows the transverse momentum-dependent factorization at small photon transverse momentum $q_{\perp}$, i.e., $q_{\perp}\ll Q$, where $Q$ defines the hard scale, representing the invariant mass of the dilepton pair. 
This factorization entails convolving perturbatively calculable hard-scattering partonic cross sections with transverse momentum-dependent parton distributions (TMDs)~\cite{Collins:2011zzd,Boussarie:2023izj,Angeles-Martinez:2015sea,Burkardt:2015qoa}. The TMDs are crucial for understanding spin and transverse momentum correlations, as well as the three-dimensional structure of hadrons in the momentum space. They are accessible in semi-inclusive deep inelastic scattering (SIDIS)~\cite{Cahn:1978se,Bacchetta:2006tn,Bacchetta:2022awv} and DY processes~\cite{Tangerman:1994eh,Boer:1997nt,Bacchetta:2019sam,Bastami:2020asv}. 

The Sivers function characterizes the asymmetric distribution of unpolarized quarks within a transversely polarized nucleon, establishing a correlation between the quark's transverse momentum and the nucleon's transverse spin~\citep{Sivers:1989cc}.
Due to its time-reversal odd (T-odd) property, the sign of the Sivers function measured in the DY process is expected to be opposite to its sign measured in SIDIS~\citep{Collins:2002kn}. In the last decade, HERMES~\citep{HERMES:2004mhh,HERMES:2009lmz}, COMPASS~\cite{COMPASS:2008isr,COMPASS:2010hbb,COMPASS:2012dmt,COMPASS:2016led}, and Jefferson Lab Hall A~\cite{JeffersonLabHallA:2011ayy} Collaborations have extensively investigated the Sivers asymmetry in SIDIS. Confirming the observed sign change is a crucial test for our understanding of QCD dynamics and the factorization scheme~\citep{Anselmino:2009st,Kang:2009bp,Peng:2014hta,Collins:2002kn,Brodsky:2002cx,Brodsky:2002rv}. This validation not only serves as a fundamental assessment but also remains a central focus on current and future DY facilities~\citep{COMPASS:2017jbv,COMPASS:2023vqt,COMPASS:2010shj,STAR:2015vmv}.  In the pion-polarized proton DY process, the convolution of the pion’s unpolarized TMD and the proton’s Sivers
functions generates a $\sin(\phi_s)$ modulated azimuthal asymmetry~\cite{Bastami:2020asv}.

Another leading twist T-odd proton distribution function, known as the Boer-Mulders function, describes the distribution of transversely polarized quarks within an unpolarized proton. It originates from the correlation between the transverse spin and transverse momentum of the quark~\citep{Boer:1999mm,Boer:1997nt}. 
In the pion-proton induced DY process, the convolution of the pion's and the proton's Boer-Mulder functions result in a $\cos(2\phi)$ azimuthal angular dependence of the final-state dilepton. 
The proton's Boer-Mulders function has been extensively studied using QCD-inspired  models~\cite{Boer:2002ju,Goldstein:2002vv,Gamberg:2007wm,Burkardt:2007xm,Bacchetta:2008af,Meissner:2007rx,Gurjar:2022rcl,Maji:2017wwd,Lyubovitskij:2022vcl,Courtoy:2009pc,Pasquini:2010af}. While the pion's Boer-Mulders function remains unexplored experimentally, various phenomenological models have been employed to calculate this distribution~\cite{Gamberg:2009uk,Pasquini:2014ppa,Wang:2017onm,Lu:2012hh,Ahmady:2019yvo,Kou:2023ady,Noguera:2015iia,Engelhardt:2015xja}.

The Kotzinian-Mulders function and pretezelosity distribution, denoted as $h_{1L}^{\perp}$ and $h_{1T}^{\perp}$, respectively describe the probability of finding a transversely polarized quark within a longitudinally and transversely polarized nucleon, respectively. Like the Boer-Mulders function, they are also chiral-odd distributions. 
In the pion-proton induced DY process, they combine with the pion Boer-Mulders function, leading to the $\sin(2\phi)$ and $\sin(2\phi+\phi_{s})$ modulated azimuthal spin asymmetry, respectively~\citep{Bastami:2020asv,Lefky:2014eia}.

In this work, we calculate four azimuthal spin asymmetries in the pion-polarized proton induced DY process: Sivers ($\sin(\phi_{s})$), $\sin(2\phi+\phi_{s})$, Boer-Mulders ($\cos(2\phi)$), and $\sin(2\phi)$ asymmetries, within the framework of TMD factorization~\cite{Collins:1984kg}. 
We utilize the leading-twist pion TMDs obtained using light-front holographic pion wave functions~\citep{Ahmady:2019yvo,Ahmady:2021lsh} and the proton TMDs calculated in a light-front quark-diquark model constructed by the soft-wall AdS/QCD~\citep{Maji:2015vsa,Gurjar:2022rcl}. We employ both the perturbative $\text{U}(1)$ and the non-perturbative $\text{SU}(3)$ gluon rescattering kernels to obtain the T-odd pion and proton TMDs. We then apply QCD
evolution of the TMDs in order to
incorporate degrees of freedom relevant to higher-resolution probes~\citep{Collins:2014jpa}. This then allows us to compare our QCD-evolved TMDs leading to the azimuthal spin asymmetries with experimental data from COMPASS Collaboration~\citep{COMPASS:2017jbv,COMPASS:2023vqt} and various phenomenological studies~\cite{Bastami:2020asv}. We further investigate the cross section for the pion-nucleus induced Drell-Yan process using the derived pion TMDs, supplemented by the TMDs of the target nucleus~\cite{Stirling:1993gc,Anassontzis:1987hk}.
%
\section{DY process with pion and polarized proton}\label{sec:Drell-Yan process with pions and polarized protons}
%
The pion-induced Drell-Yan process can be described as~\cite{Arnold:2008kf}
\begin{equation}\label{DYprocess}
	\pi^{-}(P_{\pi})+p^{\uparrow}(P_{p}) \rightarrow \gamma^{\ast}(q_{\perp})+X \rightarrow l^{+}(\ell)+l^{-}(\ell^{\prime})+X\,,
\end{equation}
where $P_{\pi}$ and $P_{p}$ denote the four-momenta of the incoming pion beam and the proton target, while $q_{\perp}$ represents the transverse momenta of the intermediate virtual photon. The momenta of the outgoing dilepton pair are designated by $\ell$ and $\ell'$. 

The total leading order differential cross section for the $\pi^{-}$p Drell-Yan process with a transversely polarized proton target is expressed by the following generic form~\cite{COMPASS:2010shj,Arnold:2008kf}
\begin{align}\label{Eq:corss_section}
	&\frac{d\sigma}{d^{4}qd\Omega}=\frac{\alpha_{\text{em}}^2}{\mathcal{F} Q^2}\Big\{\big[(1+\cos^2\theta){F_{UU}^{1}}+\sin^{2}\theta  \cos(2\phi){F_{UU}^{\cos2\phi}}\big] \nonumber\\
		&	+S_{L}\sin^{2}\theta\sin(2\phi) {F_{UL}^{\sin2\phi}}+S_{T}(1-\cos^{2}\theta)
	 \sin\phi_{S}{F_{UT}^{\sin\phi_{S}}}\nonumber\\
	&
  +S_{T}\sin^{2}\theta  \big[\sin(2\phi+\phi_{S}){F_{UT}^{\sin(2\phi+\phi_{S})}}\nonumber \\
	 &+\sin(2\phi-\phi_{S}){F_{UT}^{\sin(2\phi-\phi_{S})}}\big]\Big\}.	
\end{align}
In the above Eq.~\eqref{Eq:corss_section}, the azimuthal angle of the target polarization vector $S_{T}$ is represented as $\phi_{S}$, while in the Collins-Soper frame~\cite{Peng:2018tty}, the azimuthal and polar angles of the lepton momentum are denoted by $\phi$ and $\theta$, respectively. $\mathcal{F}$ is the flux factor of the incoming hadrons. $\Omega$ is the solid angle of the dilepton system.  
The first subscript on the structure functions $F_{XY}$ indicates that the pion is unpolarized $(U)$, while the second subscript corresponds to the proton polarization, which can be either unpolarized $(U)$, longitudinally $(L)$, or transversely $(T)$ polarized.

The single spin asymmetry is obtained by taking the ratio of polarized structure functions to unpolarized structure functions~\cite{Boussarie:2023izj,Bastami:2020asv},
\begin{align}
	A_{XY}^{\text{weight}}(x_{\pi},x_{p},q_{\perp},Q^{2})=\frac{F_{XY}^{\text{weight}}(x_{\pi},x_{p},q_{\perp},Q^{2})}{F_{UU}^{1}(x_{\pi},x_{p},q_{\perp},Q^{2})}.
\end{align}
Note that all the structure functions and asymmetries depend on the Bjorken variables $x_{\pi}$ and $x_{p}$ for the pion and proton, respectively, as well as the transverse momentum $q_{\perp}$ and virtuality $Q^{2}$ of the photon. Those structure functions can be expressed as the convolution of pion and proton TMDs~\cite{Arnold:2008kf},
	\begin{align}\label{eq:structure_functions}
		F_{UU}^{1}&=\mathcal{C}\left[f_{1,\pi}f_{1,p}\right],\nonumber\\
		F_{UU}^{\cos(2\phi)}&=\mathcal{C}\left[\frac{2({\hat{h}}.\vec{p}_{\perp\pi})({\hat{h}}.\vec{p}_{\perp p})-\vec{p}_{\perp \pi}.\vec{p}_{\perp p}}{M_{\pi}M_{p}}h_{1,\pi}^{\perp}h_{1,p}^{\perp}\right],\nonumber\\
		F_{UL}^{\sin(2\phi)}&=-\mathcal{C}\left[\frac{2({\hat{h}}.\vec{p}_{\perp\pi})({\hat{h}}.\vec{p}_{\perp p})-\vec{p}_{\perp \pi}.\vec{p}_{\perp p}}{M_{\pi}M_{p}}h_{1,\pi}^{\perp}h_{1L,p}^{\perp}\right],\nonumber\\
		F_{UT}^{\sin(\phi_{s})}&=\mathcal{C}\left[\frac{{\hat{h}}.\vec{p}_{\perp p}}{M_{p}}f_{1,\pi}f_{1T,p}^{\perp}\right],\nonumber\\
		F_{UT}^{\sin(2\phi-\phi_{s})}&=-\mathcal{C}\left[\frac{{\hat{h}}.\vec{p}_{\perp \pi}}{M_{\pi}}h_{1,\pi}^{\perp}h_{1,p}\right],\nonumber\\
		F_{UT}^{\sin(2\phi+\phi_{s})}&=-\mathcal{C}\bigg[\bigg\{\frac{2({\hat{h}}.\vec{p}_{\perp p})[2({\hat{h}}.\vec{p}_{\perp\pi})({\hat{h}}.\vec{p}_{\perp p})-\vec{p}_{\perp\pi}.\vec{p}_{\perp p}]}{2M_{\pi}M_{p}^{2}}\nonumber\\
&\hspace{1cm}-\frac{{p}_{\perp p}^{2}({\hat{h}}.\vec{p}_{\perp \pi})}{2M_{\pi}M_{p}^{2}}\bigg\}h_{1,\pi}^{\perp}h_{1T,p}^{\perp}\bigg],
	\end{align}
%
where $\mathcal{C}$ stands for the convolution integrals, ${\hat{h}}=\vec{q}_{\perp}/|q_{\perp}|$ is a unit vector pointing out along the $x$-axis in the Collins-Soper frame. The convolution integral mentioned above is defined as~\cite{Arnold:2008kf},
	\begin{align}
		&\mathcal{C}\left[\omega f_{\pi}f_{p}\right]=\sum_{i}e_{i}^{2}\int d^{2}\vec{p}_{\perp\pi}
d^{2}\vec{p}_{\perp p}\delta^{(2)}(\vec{q}_{\perp}-\vec{p}_{\perp\pi}-\vec{p}_{\perp p})\nonumber\\
&\quad\quad\quad\times\omega(\vec{q}_{\perp},\vec{p}_{\perp\pi},\vec{p}_{\perp p})f_{i,p}(x_{p},\vec{p}_{\perp p})f_{\bar{i},\pi}(x_{\pi},\vec{p}_{\perp \pi}),
	\end{align}
where the sum runs for all active quark flavors $i=u,\,\bar{u},\,d,\,\bar{d}\dots,$ and $\omega$ is a kinetic weight function, which projects out the corresponding azimuthal angular dependence. The convolution variables $\vec{p}_{\perp \pi},~\vec{p}_{\perp p}$ correspond to the relative transverse momenta of struck quarks in the pion and proton, respectively. 
%
\section{QCD evolution of structure functions}\label{sec:QCDevolutions}
To compare the model results for spin asymmetries with available experimental data, it is necessary to evolve the structure functions.
Following the TMD factorization theorem, these structure functions can be evolved from an initial model scale, $Q^{p/\pi}_{0}$, to a higher relevant scale, $Q$, using the Collins-Soper-Sterman (CSS) evolution framework~\cite{Collins:1984kg,Collins:2011zzd,Aybat:2011zv}. 
The CSS evolution of structure functions can be performed in impact parameter space.
All the evolved structure functions can be represented in terms of impact parameter-dependent TMDs and Sudakov form factors as follows~\cite{Boussarie:2023izj,Bacchetta:2019qkv},
	\begin{align}\label{eq:structurefunctions}
		F_{UU}^{1}(x_{\pi},x_{p},q_{\perp},Q^{2})&=\mathcal{B}_{0}\left[\widetilde{f}_{1,\pi}\widetilde{f}_{1,p}\right],\nonumber\\
		F_{UU}^{\cos(2\phi)}(x_{\pi},x_{p},q_{\perp},Q^{2})&=M_{\pi}M_{p}\mathcal{B}_{2}\left[\widetilde{h}_{1,\pi}^{\perp(1)}\widetilde{h}_{1,p}^{\perp(1)}\right],\nonumber\\
		F_{UL}^{\sin(2\phi)}(x_{\pi},x_{p},q_{\perp},Q^{2})&=-M_{\pi}M_{p}\mathcal{B}_{2}\left[\widetilde{h}_{1,\pi}^{\perp(1)}\widetilde{h}_{1L,p}^{\perp(1)}\right],\nonumber\\
		F_{UT}^{\sin(\phi_{s})}(x_{\pi},x_{p},q_{\perp},Q^{2})&=M_{p}\mathcal{B}_{1}\left[\widetilde{f}_{1,\pi}\widetilde{f}_{1T,p}^{\perp(1)}\right],\nonumber\\
	F_{UT}^{\sin(2\phi-\phi_{s})}(x_{\pi},x_{p},q_{\perp},Q^{2})&=-M_{\pi}\mathcal{B}_{1}\left[\widetilde{h}_{1,\pi}^{\perp(1)}\widetilde{h}_{1,p}\right],\nonumber\\
F_{UT}^{\sin(2\phi+\phi_{s})}(x_{\pi},x_{p},q_{\perp},Q^{2})&=-\frac{M_{\pi}M_{p}^{2}}{4}\mathcal{B}_{3}\left[\widetilde{h}_{1,\pi}^{\perp(1)}\widetilde{h}_{1T,p}^{\perp(2)}\right]
		\end{align}	
where $\mathcal{B}_{n}$ 
is expressed as,
\begin{equation}\label{eq:Bn}
	\begin{split}
&\mathcal{B}_{n}\left[\widetilde{F}_{\pi}\widetilde{F}_{p}\right]=\frac{1}{N_{c}}\sum_{i}e_{i}^{2}\int\frac{b_{\perp}db_{\perp}}{2\pi}b_{\perp}^{n}J_{n}(q_{\perp}b_{\perp})\\
	&\times\widetilde{F}_{\pi}^{\bar{i}}(x_{\pi},b_{\perp},Q^{\pi}_{0})\widetilde{F}_{p}^{i}(x_{p},b_{\perp},Q^{p}_{0})e^{-S(b_{\perp},Q^{\pi}_{0},Q^{p}_{0},Q)},
	\end{split}
\end{equation}
with $N_{c}$ being the number of active flavors in a particular hadron and $J_{n}$ is the Bessel function of $n$-th order. The functions $\widetilde{F}_{\pi}^{\bar{i}}$ and $\widetilde{F}_{p}^{i}$ corresponds to the pion and proton TMDs, respectively, in the impact-parameter space, obtained by performing Fourier transformation of the TMDs in momentum space at their initial model scales $Q_{0}^{\pi}$ and $Q_{0}^{p}$. The impact-parameter $b_\perp$ is the Fourier conjugate variable to $p_{\perp h}$, where index $h=\pi$ or $p$ refers to pion or proton.  The Sudakov form factor, $S(b_{\perp},Q^{\pi}_{0},Q^{p}_{0},Q)$, which contains the important effects of gluon radiation, 
is decomposed into perturbative and nonperturbative parts,
\begin{align}\nonumber
	S(b_{\perp},Q^{\pi}_{0},Q^{p}_{0},Q)=&S_{\textrm{pert}}(b_{\ast},Q)+S_{\textrm{NP}}^{\rm p}(b_{\perp},Q^{p}_{0},Q)\nonumber\\&+S_{\textrm{NP}}^{\rm \pi}(b_{\perp},Q^{\pi}_{0},Q).
\end{align} 
The perturbative part is given by~\cite{Collins:2011zzd},
\begin{align}\label{eq:Spert}
S_{\textrm{pert}}=\int^{Q}_{Q_i}\frac{d\bar{\mu}}{\bar{\mu}}\left[A\left(\alpha_s(\bar{\mu})\right)
\mathrm{ln}\frac{Q^2}{\bar{\mu}^2}+B(\alpha_s(\bar{\mu}))\right],	
\end{align}
where $Q_{i}=2e^{-\gamma_{E}}/b_{\ast}$ (with $\gamma_{E}\simeq0.577$), 
with  the  choice of $b_{\ast}$ in such a way that $b_{\ast}(\bp)=\bp/({1+\frac{\bp^{2}}{b_{\textrm{max}}^{2}}})^{1/2} \simeq{b_{\textrm{max}}}$  at $\bp\rightarrow\infty$ and $b_{\ast}(\bp)\simeq \bp$ when $\bp \rightarrow 0$~\cite{Collins:2014jpa}. These allow avoiding the Landau pole by freezing the scale $\bp$~\cite{Collins:1984kg}. $b_{\text{max}}$ separates the perturbative and nonperturbative domains of the TMDs and it is determined phenomenologically as $b_{\textrm{max}}=1.5$ GeV$^{-1}$~\cite{Sun:2014dqm}.
The above perturbative Sudakov Form factor is spin-independent and has the same form for all kind of distribution functions. The coefficients $A$ and $B$ in Eq.~(\ref{eq:Spert}) are the anomalous dimensions and can be expanded perturbatively~\cite{Collins:2011zzd,Aybat:2011ge,Echevarria:2014xaa}.

On the other hand, the nonperturbative proton and pion Sudakov form factors, $S_{\text{NP}}^{\rm p/\pi}$, have been studied phenomenologically~\cite{Vladimirov:2019bfa,Cerutti:2022lmb}. A generic form for $S_{\rm{NP}}$ proposed in Ref.~\cite{Collins:1984kg} is given by,
\begin{equation}
	\label{eq:snp_gene}
	S_{\rm{NP}}^{\rm p/\pi}(\bp,Q^{\rm p/\pi}_{0},Q)=g_1^{\rm p/\pi}(\bp)+g_2^{\rm p/\pi}(\bp)\ln \frac{Q}{Q^{\rm p/\pi}_0},
\end{equation}
where $g^{p/\pi}_{1}(\bp)$ and $g^{p/\pi}_{2}(\bp)$ depend on the hadronic distribution functions. For the proton, they read as
\begin{eqnarray}
	g_{1}^{p}(\bp)=\frac{g_{1}^{p}}{2}\bp^{2},~~~g_{2}^{p}(\bp)=\frac{g_{2}^{p}}{2}\ln\frac{\bp}{b_{\ast}}
\end{eqnarray}
with $g_{1}^{p}=0.212 \pm 0.006
$ GeV$^2$ and $g_{2}^{p}= 0.84 \pm 0.037
$ GeV$^2$~\cite{CDF:2012brb,D0:2007lmg}, whereas for the pion, they are given by,
\begin{eqnarray}
	g_{1}^{\pi}(\bp)={g_{1}^{\pi}}\bp^{2},~~~g_{2}^{\pi}(\bp)=g_{2}^{\pi}\ln\frac{\bp}{b_{\ast}}.
\end{eqnarray}
The numerical values of $g_1^\pi$ and $g_2^\pi$ are obtained by fitting to the $\pi^- N$ DY data~\cite{Conway:1989fs}: 
$g^{\pi}_{1}=0.082 \pm 0.022
$ GeV$^2$ and $g^{\pi}_{2}=0.394 \pm 0.103 
$ GeV$^2$. By employing Eqs.~(\ref{eq:structurefunctions}) and (\ref{eq:Bn}), the QCD evolution of both unpolarized and polarized structure functions can be performed from the initial hadronic model scale $Q^{p/\pi}_{0}$ to the relevant experimental final scale $Q$. 

%
\section{Model inputs for the pion and proton TMDs}\label{sec:model_inputs}
\subsection{Pion TMDs}
We employ the holographic light-front pion wave functions for the pion TMDs~\cite{Brodsky:2014yha,Bacchetta:2017vzh}. For the pion, there exist two leading-twist TMDs: the unpolarized quark TMD, $f^q_{1,\pi}(x,k_\perp^2)$, and the Boer-Mulders function, $h_{1,\pi}^{\perp q}(x,k_\perp^2)$~\cite{Pasquini:2014ppa}. The pion unpolarized TMD can be expressed in terms of the light-front wave functions (LFWFs) of pion as,
\begin{equation}
 	f_{1,\pi}^{q}(x,k_\perp^2)=\frac{1}{16\pi^3} \sum_{h,\bar{h}} |\Psi_{h \bar{h}}(x,\vec{k}_\perp)|^2  \;,
 \label{hf1}
 \end{equation} 
 where $\Psi_{h \bar{h}}(x,\vec{k}_\perp)$ represents the pion's LFWF with $h\, (\bar{h})$ being the helicity of the quark (antiquark) in the leading Fock sector. The normalization condition for the unpolarized TMD is as follows:
 \begin{equation}
 	\int \mathrm{d} x \,\mathrm{d}^2 \vec{k}_\perp \,f_{1,\pi}^{q}(x,k_\perp^2)=1 .
 \label{PDF-norm}
 \end{equation}

To obtain the nonzero pion Boer-Mulders function $h_{1,\pi}^{\perp q}(x,{k}_\perp^2)$, it is necessary to account for the gauge link. This represents the initial or final-state interactions of the active parton with the target remnant, collectively referred to as the gluon rescattering kernel. The pion Boer-Mulders function can be represented as follows~\cite{Ahmady:2019yvo}:
 \begin{align}
&	k_\perp^2 h_{1,\pi}^{\perp q}(x,k_\perp^2) = M_\pi \int \frac{\mathrm{d}^2 \vec{k}_\perp^{\prime}}{16\pi^3}~ i G(x, |\vec{k}_\perp-\vec{k}_\perp^{\prime}|) \nonumber\\
	&\times \sum_{h,\bar{h}} \Psi_{-h,\bar{h}}^*(x, \vec{k}^{\prime}_\perp) h k_\perp e^{i h \theta_{k_\perp}} \Psi_{h,\bar{h}}(x,\vec{k}_\perp) \;,
\label{BM-overlap}
\end{align}
where $G(x, |\vec{k}_\perp-\vec{k}_\perp^{\prime}|)$ corresponds to the gluon rescattering kernel with $(\vec{k}_\perp-\vec{k}_\perp^{\prime})$ being the transverse momentum carried by the exchanged gluon.  $\theta_{k_\perp}$ is the polar angle in a two-dimensional polar coordinate system, ranging from 0 to $2\pi$. To proceed further, we must specify the form of the gluon rescattering kernel. A straightforward approach is to assume the perturbative Abelian gluon rescattering kernel, as defined by~\cite{Bacchetta:2008af,Wang:2017onm}, 
\begin{equation}
 	i G^{\mathrm{pert.}}(x, {|\vec{k}_\perp-\vec{k}_\perp^{\prime}|}) = \frac{C_F\alpha_s}{2\pi}\frac{1}{(|\vec{k}_\perp-\vec{k}_\perp^{\prime}|)^2} \;,
 	\label{pert-G}
\end{equation}
with the color factor $C_F=4/3$ and $\alpha_s$ being the fixed coupling constant. An exact computation of the nonperturbative gluon rescattering kernel is currently unavailable. In practice, an approximation scheme is necessary. Meanwhile, the gluon rescattering kernel can be expressed in terms of the QCD lensing function $I(x, |\vec{k}_\perp-\vec{k}_\perp^{\prime}|)$, as described in Ref.~\cite{Ahmady:2019yvo},
\begin{equation}
i G(x, |\vec{k}_\perp-\vec{k}_\perp^{\prime}|)= -\frac{2}{(2\pi)^2} \frac{(1-x) I(x, |\vec{k}_\perp-\vec{k}_\perp^{\prime}|)}{(|\vec{k}_\perp-\vec{k}_\perp^{\prime}|)} \;.
	\label{Relation-GI}
\end{equation}
The above relation originates from the connection between the chiral-odd GPD and the first moment of the pion Boer-Mulders function~\cite{Burkardt:2007xm}. Gamberg and Schlegel deduced the QCD lensing function $I(x, |\vec{k}_\perp-\vec{k}_\perp^{\prime}|)$ from the eikonal amplitude for final-state rescattering through the exchange of non-Abelian SU$(3)$ soft gluons~\cite{Gamberg:2009uk}. The nonperturbative gluon rescattering kernel, Eq.~\eqref{Relation-GI}, has been effectively used to calculate T-odd TMDs for both the pion~\cite{Ahmady:2019yvo,Kou:2023ady} and the proton~\cite{Gurjar:2022rcl}.

To compute the pion's TMDs, we utilize the spin-improved holographic wave function~\cite{Ahmady:2018muv,Ahmady:2019yvo}, 
	\begin{align}
	 	\Psi_{h,\bar{h}}(x,\vec{k}_\perp)=& \Big[ (M_{\pi} x(1-x) + B m_q) h\delta_{h,-\bar{h}}  \nonumber\\
	 	&- B    k_\perp e^{-ih\theta_{k_\perp}}\delta_{h,\bar{h}}	\Big] \frac{\Psi (x, \vec{k}_\perp)}{x(1-x)}\,.
	 \label{spin-improved-wfn-k}
	 \end{align}
	 where 
	 	 \begin{equation}
	 	\Psi (x,\vec{k}_\perp)=\mathcal{N} \frac{1}{\sqrt{x (1-x)}}  \exp{ \Big[ -\frac{k_\perp^2+m_q^2}{2\kappa^2 x(1-x)} \Big] } \,,
 \label{hWF-k}
 \end{equation}
 with $m_q$ being the effective quark mass and $\mathcal{N}$ is a normalization constant determined using
 \begin{equation}
 	\sum_{h,\bar{h}}\int \mathrm{d} x \frac{\mathrm{d}^2 \vec{k}_\perp}{16\pi^3} |\Psi_{h \bar{h}}(x,\vec{k}_\perp)|^2 =1 \;.
\end{equation} 

The parameter $B$ in Eq.~\eqref{spin-improved-wfn-k} is known as the dynamical spin parameter. When $B \to 0$, it signifies no spin-orbit correlations, akin to the original holographic wave function~\cite{Brodsky:2014yha}. Conversely, when $B \geq 1$, it indicates maximal spin-orbit correlation. 

With $B \geq 1$, the constituent quark mass $m_{q}=330$ MeV~\cite{Ahmady:2019yvo}, and AdS/QCD scale parameter $\kappa=523\pm 24$ MeV determined by simultaneous fit to the Regge slopes of mesons and baryons~\cite{Brodsky:2016yod}, the pion wave function has been effectively utilized to calculate various pion observables, including electromagnetic form factors, associated radii, transition form factor, parton distribution function (PDF), TMDs, etc., with notable overall success~\cite{Ahmady:2016ufq,Ahmady:2018muv,Ahmady:2019yvo,Gurjar:2023uho,Ahmady:2020mht}.

Note that in Eq.~\eqref{hWF-k}, the longitudinal mode is not dynamically generated. Brodsky and de Téramond proposed this prescription based on the invariant mass ansatz (IMA) to describe the longitudinal mode~\cite{Brodsky:2008pg}. In contrast, Refs.~\cite{Ahmady:2021lsh,Ahmady:2021yzh} incorporate longitudinal dynamics generated by the 't Hooft equation to describe the full meson spectrum, with specific pion dynamics predicted in Ref.~\cite{Ahmady:2022dfv}. The idea of extending beyond the invariant mass prescription using the 't Hooft equation was initially suggested in Ref.~\cite{Chabysheva:2012fe} to forecast meson decay constants and parton distribution functions. Recently, Refs.~\cite{deTeramond:2021yyi,Li:2021jqb} surpassed this prescription by implementing a phenomenological longitudinal confinement potential, initially introduced in Ref.~\cite{Li:2015zda} within the framework of basis light-front quantization. While Refs.~\cite{deTeramond:2021yyi,Li:2021jqb} focus on the chiral limit and chiral symmetry breaking. Ref.~\cite{deTeramond:2021yyi} extends its exploration to ground state heavy mesons and investigates connections with the 't Hooft equation.

It is worth noting that there has been a notable increase in interest in incorporating longitudinal dynamics within holographic light-front QCD (hLFQCD)~\cite{Li:2021jqb,deTeramond:2021yyi,Lyubovitskij:2022rod,Weller:2021wog,Ahmady:2021lsh,Rinaldi:2022dyh}. The resulting meson wave function with 't Hooft longitudinal mode can be expressed as~\cite{Ahmady:2022dfv,Gurjar:2024wpq}
\begin{align}
    \Psi (x,\vec{k}_\perp) = \mathcal{N} \frac{1}{\sqrt{x (1-x)}} \chi_{\rm tH}(x) \exp{ \left[-\frac{k_\perp^2}{2\kappa^2 x(1-x)} \right] } \,,
    \label{hWF-ktHooft}
\end{align}
where the numerical solutions for the longitudinal modes $\chi_{\rm tH}(x)$ of the ground state meson can be approximately fitted to the following polynomial form:
\begin{align}
    \chi_{\rm tH}(x) \simeq x^{\beta_{1}} (1 - x)^{\beta_{2}} ,
    \label{eq:tHooftanaly}
\end{align}
with $\beta_i$ being quark mass-dependent variables, which vanish in the chiral limit. For the ground state of light unflavored mesons, we find $\beta_{1,2} = 0.51$ with $m_q = 46$ MeV~\cite{Ahmady:2022dfv,Gurjar:2024wpq}.

Using the pion wave function with IMA provided in {Eqs.~\eqref{spin-improved-wfn-k} and \eqref{hWF-k}}, the explicit expression for the unpolarized quark TMD is given as follows:
	\begin{align}\label{eq:pionunpolTMD}
		f_{1,\pi}^{q}(x,\kp^2)&=\frac{2 \mathcal{N}^{2}}{16\pi^{3}}\frac{(M_{\pi}x(1-x)+Bm_{q})^{2}+B^{2}\kp^{2}}{(x(1-x))^{3}}\nonumber\\&\times \exp\Big[-\frac{\kp^{2}+m_{q}^{2}}{x(1-x)\kappa^{2}}\Big]\,.
  \end{align}
 {Meanwhile, the pion unpolarized TMD with the 't Hooft longitudinal mode can be derived from the wave function in Eqs.~\eqref{spin-improved-wfn-k} and \eqref{hWF-ktHooft} as:
 \begin{align}
    f_{1,\pi}^{q}(x,\kp^2)&=\frac{2 \mathcal{N}^{2}}{16\pi^{3}}\frac{(M_{\pi}x(1-x)+Bm_{q})^{2}+B^{2}\kp^{2}}{(x(1-x))^{3}}\nonumber\\&\times x^{2\beta_{1}}(1-x)^{2\beta_{2}}\exp\Big[-\frac{\kp^{2}}{x(1-x)\kappa^{2}}\Big]\,. 
 \end{align}}
  Using the perturbative gluon rescattering kernel given in Eq.~\eqref{pert-G}, we derive an analytical expression for the pion Boer-Mulders function {for IMA} as, 
  	\begin{align}\label{eq:pionBMTMD}
&		h_{1,\pi}^{\perp q}(x,\kp^{2})=\alpha_{s}B C_{F}\frac{M_{\pi}\mathcal{N}^{2}}{4\pi^{3}}\frac{M_{\pi}x(1-x)+Bm_{q}}{(x(1-x))^{2}}\Big(\frac{\kappa}{\kp}\Big)^{2}\nonumber \\&\times\exp\Big[-\frac{\kp^{2}+2m_{q}^{2}}{2\kappa^{2}x(1-x)}\Big]\Big(1-\exp\Big[-\frac{\kp^{2}}{2\kappa^{2}x(1-x)}\Big]\Big)\,.
	\end{align}
 {With 't Hooft longitudinal mode, the pion Boer-Mulders function reads as, 
 \begin{align}
	h_{1,\pi}^{\perp q}(x,\kp^{2})=&\alpha_{s}B C_{F}\frac{M_{\pi}\mathcal{N}^{2}}{4\pi^{3}}\frac{M_{\pi}x(1-x)+Bm_{q}}{(x(1-x))^{2}}\Big(\frac{\kappa}{\kp}\Big)^{2}\nonumber \\&\times x^{2\beta_{1}}(1-x)^{2\beta_{2}}\exp\Big[-\frac{\kp^{2}}{2\kappa^{2}x(1-x)}\Big]\nonumber \\&\times\Big(1-\exp\Big[-\frac{\kp^{2}}{2\kappa^{2}x(1-x)}\Big]\Big)\,.
	\end{align}}
If $B\to 0$, the holographic Boer-Mulders function becomes zero. However, for $B \ge 1$, it is hardly affected by the parameter $B$, as the normalization constant $\mathcal{N}$ of the wave function scales as $1/B^2$. Since, the nonperturbative gluon rescattering kernels~\cite{Ahmady:2019yvo} do not allow for an analytical expression of the pion Boer-Mulders function, we proceed with its numerical computation. The standard deviation for $\kappa$ is around 5\%. Similarly, we apply a 5\% uncertainty in the quark mass~\cite{Gurjar:2023uho}.

\subsection{Proton TMDs}
 For the proton TMDs, we utilize the light-front quark-diquark model constructed by the soft-wall AdS/QCD~\cite{Gutsche:2013zia,Chakrabarti:2015lba,Chakrabarti:2015ama,Chakrabarti:2016yuw}. The T-even TMDs of the proton relevant to the structure functions, Eq.~\eqref{eq:structure_functions}, can be expressed in terms of the LFWFs as follows:
 \begin{align}\label{eq:unpol_proton}
     f_{1,p}^{q}(x,k_{\perp}^{2})=&\frac{1}{16\pi^{3}}\Big[|\Psi_{+}^{+}(x,\vec{k}_\perp)|^{2}+|\Psi_{-}^{+}(x,\vec{k}_\perp)|^{2}\Big],\\
       h_{1,p}^{q}(x,\kp^{2})=&\frac{1}{16\pi^{3}}\Big[\Psi^{+\ast}_{+}(x,\vec{k}_\perp)\Psi^{-}_{-}(x,\vec{k}_\perp)\Big],
     \\
       h_{1L,p}^{\perp q}(x,\kp^{2})=&\frac{1}{16\pi^{3}}\frac{2M}{\kp^{2}}\Big[k_\perp^r\Psi^{+\ast}_{-}(x,\vec{k}_\perp)\Psi^{+}_{+}(x,\vec{k}_\perp)\Big],\\
       h_{1T,p}^{\perp q}(x,\kp^{2})=&\frac{1}{16\pi^{3}}\frac{M^2}{k_1^2-k_2^2}\Big[\Psi^{+\ast}_{-}(x,\vec{k}_\perp)\Psi^{-}_{+}(x,\vec{k}_\perp) \nonumber\\
     &+\Psi^{-\ast}_{+}(x,\vec{k}_\perp)\Psi^{+}_{-}(x,\vec{k}_\perp)\Big],
 \end{align}
 where $\Psi^{\lambda_{N}}_{\lambda_{q}}(x,\vec{k}_\perp)$ are the LFWFs with the proton
helicities $\lambda_{N}=\pm$ and for the quark $\lambda_{q}=\pm$; plus and minus correspond to $+\frac{1}{2}$ and $-\frac{1}{2}$, respectively and $k_\perp^{r(l)}=k_1\pm ik_2$.

For the T-odd TMDs, the Sivers and Boer-Mulders functions, it is essential to consider the gauge link. Using  the gluon rescattering kernel, $G(x, |\vec{k}_\perp-\vec{k}_\perp^{\prime}|)$, and the LFWFs, they can be represented as,
 \begin{align}\nonumber
     f_{1T,p}^{\perp q}(x,\kp^{2})=&\frac{2M}{k_{\perp}^l}\int\frac{{\rm d}^{2}\vec{k}_\perp^{\prime}}{16\pi^{3}}  iG(x,|\vec{k}_\perp-\vec{k}_\perp^{\prime}|)\nonumber\\
     &\times\sum_{\lambda_{q}}\Big[\Psi^{+\ast}_{\lambda_{q}}(x,\vec{k}_\perp^{\prime})\Psi^{-}_{\lambda_{q}}(x,\vec{k}_\perp)\Big],\\
      h_{1,p}^{\perp q}(x,\kp^{2})=&\frac{2M}{k_{\perp}^l}\int\frac{{\rm d}^{2}\vec{k}_\perp^{\prime}}{16\pi^{3}}iG(x,|\vec{k}_\perp-\vec{k}_\perp^{\prime}|)\nonumber\\
      &\times\sum_{\lambda_{q},\lambda_{q}^{\prime}}\Big[\Psi^{+\ast}_{\lambda_{q}^{\prime}}(x,\vec{k}_\perp^{\prime})
 \Psi^{+}_{\lambda_{q}}(x,\vec{k}_\perp)\Big].
 \label{eq:proton_BM_TMD}
 \end{align}

 \begin{figure*}
    \centering
    \includegraphics[width=\linewidth]{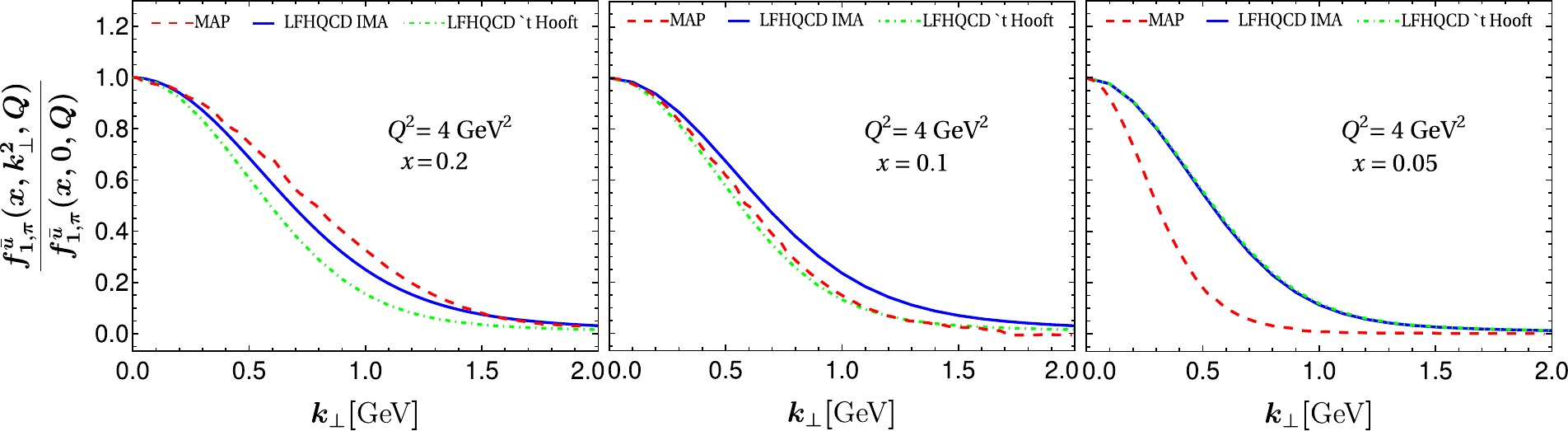}
    \caption{{Comparison of model results for the IMA and 't Hooft pion unpolarized TMDs with phenomenological extractions from the MAP collaboration at three different values of longitudinal momentum fractions, $x = 0.2$, $x = 0.1$, and $x = 0.05$, from the left to right panels, respectively.} }
    \label{fig:TMDs_comparison_pion}
\end{figure*}
\begin{figure*}
    \centering
    \includegraphics[width=\linewidth]{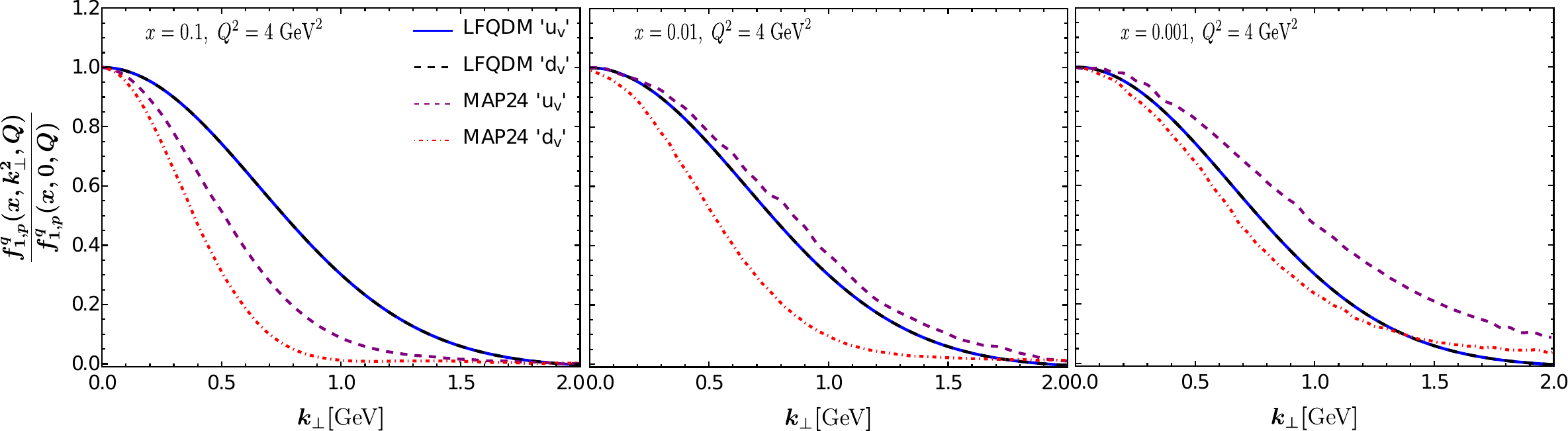}
    \caption{{Comparison of normalized unpolarized valence quark proton TMDs between our model predictions and the extracted results in the MAPTMD24 fit from MAP collaborations for three different values of longitudinal momentum fractions, $x=0.1,~x=0.01$ and $x=0.001$ at $Q^{2}= 4~\text{GeV}^{2}$. }}
    \label{fig:TMDs_comparison_b}
\end{figure*}

Using the proton LFWFs from the quark-diquark model~\cite{Chakrabarti:2015lba,Chakrabarti:2015ama,Chakrabarti:2016yuw}, one can derive the analytical expressions for the quark TMDs by substituting them into the above Eqs.~\eqref{eq:unpol_proton}-\eqref{eq:proton_BM_TMD}. The expressions for the T-even TMDs read as
 \begin{align}
 \label{unpoltmd}
f_{1,p}^{q}\left(x,\kp^2 \right)=&\frac{\log (1 / x)}{\pi \kappa^{2}}\Big(F_{1}^{q}(x)+\frac{\kp^{2}}{M^{2}} F_{2}^{q}(x)\Big)\nonumber\\
&\times \exp \Big[-\frac{\kp^{2} \log (1 / x)}{\kappa^{2}(1-x)^{2}}\Big],\\
h_{1,p}^{q}(x,\kp^2)=&\frac{\text{log}(1/x)}{\pi\kappa^{2}}F_{1}^{q}(x)\exp \Big[-\frac{\kp^{2} \log (1 / x)}{\kappa^{2}(1-x)^{2}}\Big],\\
 	h_{1L,p}^{\perp q}(x,\kp^2)=&-\frac{2\text{log}(1/x)}{\pi\kappa^{2}}F_{3}^{q}(x)\exp \Big[-\frac{\kp^{2} \log (1 / x)}{\kappa^{2}(1-x)^{2}}\Big],	\\
  h_{1T,p}^{\perp q}(x,\kp^2)=&-\frac{2\text{log}(1/x)}{\pi\kappa^{2}}F_{2}^{q}(x)\exp \Big[-\frac{\kp^{2} \log (1 / x)}{\kappa^{2}(1-x)^{2}}\Big],
  \end{align}
where the parametrized functions $F_{i}^{q}(x)$ are given by,
\begin{align}\label{F12x}
	F_{1}^{q}(x)&=\left|N_{q}^{(1)}\right|^{2} x^{2 a_{q}^{(1)}}(1-x)^{2 b_{q}^{(1)}-1} \,,\nonumber  \\
	F_{2}^{q}(x)&=\left|N_{q}^{(2)}\right|^{2} x^{2 a_{q}^{(2)}-2}(1-x)^{2 b_{q}^{(2)}-1}\,,\nonumber  \\
	F_{3}^{q}(x)&=N_{q}^{(1)}N_{q}^{(2)}x^{a_{q}^{(1)}+a_{q}^{(2)}-1}(1-x)^{b_{q}^{(1)}+b_{q}^{(2)}-1}.
\end{align}
The model parameters $a_{q}^{i}$, $b_{q}^{i}$ and the normalization constants $N_q^i$, along with the AdS/QCD scale parameter $\kappa$ are determined by fitting the electromagnetic properties of the nucleons~\cite{Mondal:2016afg,Chakrabarti:2015ama}. This model has been widely used to study and effectively reproduce various noteworthy properties of the proton~\cite{Chakrabarti:2016yuw,Chakrabarti:2015ama,Chakrabarti:2015lba,Mondal:2017wbf,Maji:2015vsa,Choudhary:2022den,Gurjar:2022rcl,Mondal:2016afg}.

 Using the perturbative gluon rescattering kernel, Eq.~\eqref{pert-G}, the expressions for the Sivers and Boer-Mulders functions read as, 
 \begin{align}\label{eq:siverstmd}\nonumber
	&f_{1 T,p}^{\perp q}(x,\kp^2)\equiv  h_{1(p)}^{\perp q}(x,\kp^2)=\frac{4}{\pi}C_{F}\alpha_{s}(1-x)^{2} F_{3}^{q}(x)\frac{1}{\kp^{2}} \nonumber\\
	 & \times \exp\Big[-\frac{\kp^{2}\log(1/x)}{\kappa^{2}(1-x)^{2}}\Big]\Big(1-\exp\Big[\frac{\kp^{2}\log(1/x)}{2\kappa^{2}(1-x)^{2}}\Big]\Big)\,.
\end{align}
In our scalar quark-diquark model, both the Sivers and Boer-Mulders TMDs for up and down quarks have the same sign, a distinct feature of this model as seen in prior study~\cite{Lyubovitskij:2022vcl,Bacchetta:2008af,Hwang:2010dd,Boer:2002ju}. Introducing the axial-vector diquark may alter these relations for the up and the down quarks~\cite{Maji:2017wwd,Bacchetta:2008af,Ellis:2008in,Gurjar:2021dyv}. In a similar fashion to the pion's Boer-Mulders TMD, we conduct numerical computations for the proton's T-odd TMDs employing nonperturbative gluon rescattering kernels.

\begin{figure*}[ht]
	\centering
	\includegraphics[width=\textwidth]{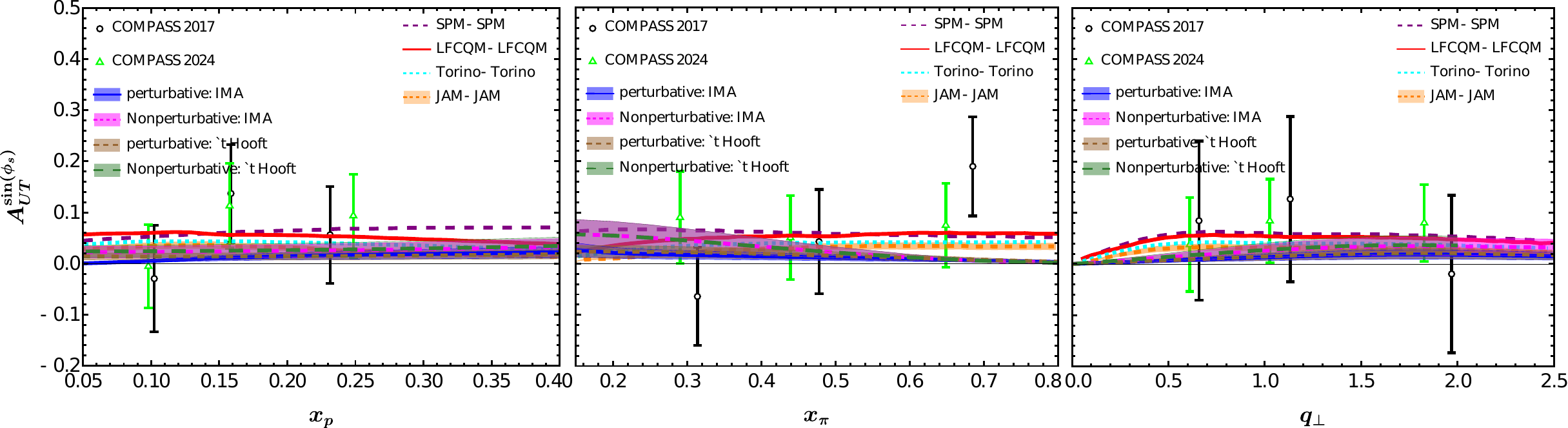}
	\caption{$\sin(\phi_{S})$ azimuthal asymmetry in pion-proton induced DY process. The panels from left to right show the variation of the azimuthal asymmetry with $x_{p}$, $x_{\pi}$, and $q_{\perp}$, respectively. The black open circles and green up-triangles represent the COMPASS data~\cite{COMPASS:2017jbv,COMPASS:2023vqt}. Our estimations (blue and magenta bands {are for IMA, while brown and light-green bands are for 't Hooft}) are compared with the results obtained from pure model and hybrid calculations reported in Ref.~\cite{Bastami:2020asv}. The purple-dashed and solid-red lines show the pure model calculations from spectator model (SPM)~\cite{Gamberg:2009uk} and light-front constituent quark model (LFCQM)~\cite{Pasquini:2014ppa}, while the orange band and cyan dashed lines represent the phenomenological computations from JLab angular momentum (JAM)~\cite{Cammarota:2020qcw} and Torino~\cite{Anselmino:2011gs} collaborations, respectively.}
 \label{fig:sin(phi)} 
\end{figure*}
\begin{figure*}[ht]
	\centering
	\includegraphics[width=\textwidth]{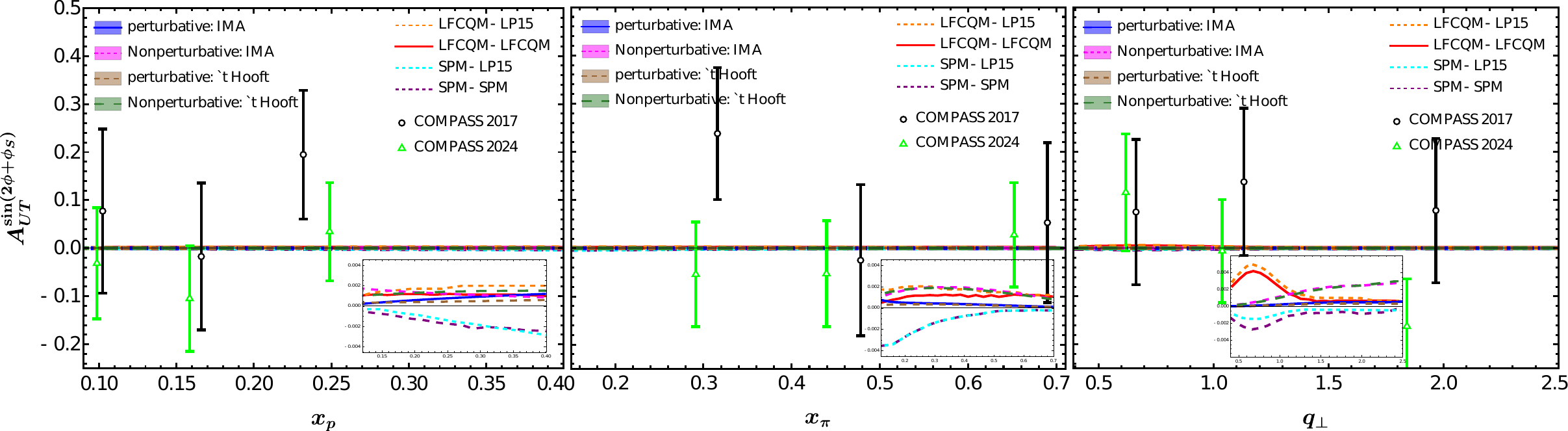}
	\caption{$\sin(2\phi+\phi_{S})$ azimuthal asymmetry in pion-proton induced DY process. The panels from left to right show the variation of the azimuthal asymmetry with $x_{p}$, $x_{\pi}$, and $q_{\perp}$, 
 respectively. The black open circles and green up-triangles represent the COMPASS data~\cite{COMPASS:2017jbv,COMPASS:2023vqt}. Our estimations (blue and magenta bands {are for IMA, while brown and light-green bands are for 't Hooft}) are compared with the results obtained from pure model and hybrid calculations reported in Ref.~\cite{Bastami:2020asv}. The orange and cyan lines represent the hybrid computations of SPM and LFCQM models with LP15 extractions of pretzelosity TMD~\cite{Lefky:2014eia}, while the purple-dashed and solid red lines shows the pure model calculations from SPM~\cite{Gamberg:2009uk} and LFCQM~\cite{Pasquini:2014ppa} models, respectively.}
	\label{fig:sin(2phi+phis)} 
\end{figure*}
%
%
\section{Numerical results}\label{sec:results}
{Figure~\ref{fig:TMDs_comparison_pion} presents a comparison between our model predictions for the normalized unpolarized pion TMDs and recent phenomenological extractions from experimental data by the MAP Collaboration~\cite{Cerutti:2022lmb}. The left panel shows the variation with $|\bfk|$ at $x = 0.2$, incorporating the evolution effects of the TMDs at $Q^{2}=4~\text{GeV}^{2}$. The middle and right panels correspond to $x = 0.1$ and $x = 0.05$, respectively. It is important to note that the MAP results include contributions from all quark flavors, i.e., both valence and sea quarks, while our model results associated only with to valence quark distributions. As evident from the figure, our model predictions agree well with the MAP results at large values of $x$, where valence quarks dominate. However, at lower values of $x$, where sea quark contributions become significant, our model predictions deviate from the MAP results. The MAP results decrease sharply compared to our model predictions.}

{Figure~\ref{fig:TMDs_comparison_b} 
presents a comparison of the normalized unpolarized proton TMDs for valence quarks with MAP extractions~\citep{Bacchetta:2024qre}. The panels, from left to right, correspond to longitudinal momentum fractions of $x = 0.1$, $0.01$, and $0.001$, respectively. We find that our model predictions for the normalized TMDs of up and down quarks overlap with each other, while the extracted TMDs from MAP show distinct differences. Additionally, for large longitudinal momentum fractions, the extracted results drop sharply compared to our model predictions, whereas for smaller momentum fractions, our model predictions are consistent with the MAP results.}

{Employing the model results of pion and proton TMDs}, we analyze the single azimuthal transverse spin asymmetries in the polarized DY process within the kinematics of the COMPASS experiment, where the DY cross section for a proton target is predominantly influenced by contributions from the proton's $u$-quark and the pion's $\bar{u}$-quark TMDs~\cite{COMPASS:2017jbv},
\begin{equation}
\begin{split}
	&0.05<x_N<0.4,~0.05<x_\pi<0.9,~ -0.3<x_F<1\\
	&
	4.3\ \mathrm{GeV}<Q<8.5\ \mathrm{GeV},
	\quad s=357~\mathrm{GeV}^2.
	\label{eq:kinematics}
	\end{split}
\end{equation}
 We evolve the pion TMDs from the
model scale $Q_{0}^{\pi}\sim 0.316$ GeV~\cite{Ahmady:2018muv,Gurjar:2023uho} to the scale $Q_f\sim 6.4$ GeV relevant to the
experimental data for the asymmetries following QCD
evolution discussed in section~\ref{sec:QCDevolutions}. The proton TMDs are also evolved from the model scale  of the quark-diquark model $Q_0^{p}\sim 0.56$ GeV~\cite{Gurjar:2022rcl,Choudhary:2022den} to the  experimental scale.

Figure~\ref{fig:sin(phi)} presents the $\sin(\phi_{s})$ asymmetry, which is known as the Sivers asymmetry. This asymmetry arises from the convolution of the unpolarized parton distribution of the incoming pion beam and the Sivers function of the transversely polarized target proton, i.e., $A_{UT}^{\sin(\phi_{s})}\propto f_{1,\pi}^{\bar{q}}\otimes f_{1T,p}^{\perp q}$. The perturbative and nonperturbative curves represent the asymmetries generated by using the perturbative and nonperturbative
gluon rescattering kernels for the proton Sivers function. We compare our pure model results with recently reported COMPASS data~\cite{COMPASS:2017jbv,COMPASS:2023vqt} and also with available theoretical predictions reported in Ref.~\cite{Bastami:2020asv}, where the nonperturbative input for the pion  TMD is taken from the light-front constituent quark model (LFCQM)~\cite{Pasquini:2014ppa} and the spectator model (SPM)~\cite{Gamberg:2009uk}, and the proton TMD is adopted from the LFCQM~\cite{Pasquini:2008ax,Boffi:2009sh,Pasquini:2011tk} and the SPM~\cite{Gamberg:2007wm} as well as from the available Parametrization of TMDs extracted from the experimental data by Torino Collaboration~\cite{Anselmino:2013vqa} and JAM20 Collaboration~\cite{Cammarota:2020qcw}. 
We observe that our model computations for the Sivers asymmetry are consistently positive across the entire range of COMPASS kinematics, which aligns well with both the COMPASS 2017 and COMPASS 2024 results~\cite{COMPASS:2017jbv,COMPASS:2023vqt}. Disregarding the contribution from sea quarks, we can infer that $A_{UT}^{\sin(\phi_{s})}\propto f_{1T,p}^{\perp u}>0$ as the unpolarized distribution of the pion's valence quarks is always positive~\cite{Cerutti:2022lmb}. This suggests that the $u$-quark Sivers function of the proton is positive in the DY process, whereas it is negative in the SIDIS process, a finding consistent with various theoretical analyses~\cite{Pasquini:2011tk,Gurjar:2022rcl,Maji:2017wwd,Bacchetta:2008af} and phenomenological studies~\cite{Efremov:2004tp,Anselmino:2009st,Anselmino:2008sga,Anselmino:2005ea,Fernando:2023obn,Bury:2021sue,Bury:2020vhj}. The Sivers asymmetry is unique in that it can be fully described by both model predictions and available parametrizations like JAM20~\cite{Cammarota:2020qcw} and Torino~\cite{Anselmino:2011gs}. It serves as confirmation of the sign change of the Sivers function in both SIDIS (final state interactions) and DY (initial state interactions) processes observed in COMPASS~\cite{COMPASS:2017jbv,COMPASS:2023vqt} and RHIC~\cite{Aschenauer:2015eha}.

\begin{figure*}[ht]
	\centering
    \includegraphics[width=\textwidth]{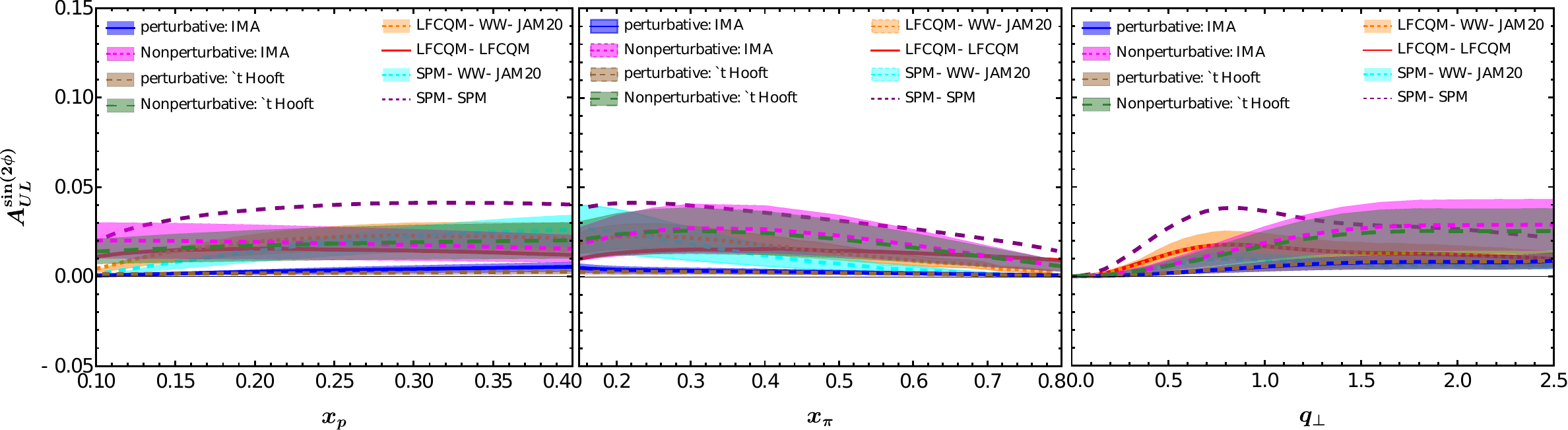}
	\caption{$\sin(2\phi)$ azimuthal asymmetry in pion-proton induced DY process. The panels from left to right show the variation of the azimuthal asymmetry with $x_{p}$, $x_{\pi}$, and $q_{\perp}$, 
 respectively. Our estimations (blue and magenta bands {are for IMA while brown and light-green bands are for 't Hooft}) are compared with the results obtained from pure model and hybrid calculations reported in Ref.~\cite{Bastami:2020asv}. The orange and cyan bands represent the hybrid computations of LFCQM and SPM models with JAM20 fit of proton WW-type approximation of Kotzinian-Mulders function~\cite{Bastami:2018xqd}, while the purple-dashed and solid-red lines shows the pure model calculations from SPM~\cite{Gamberg:2009uk} and LFCQM~\cite{Pasquini:2014ppa} models, respectively.}
	\label{fig:sin(2phi)} 
\end{figure*}
\begin{figure*}[ht]
	\centering
	\includegraphics[width=\textwidth]{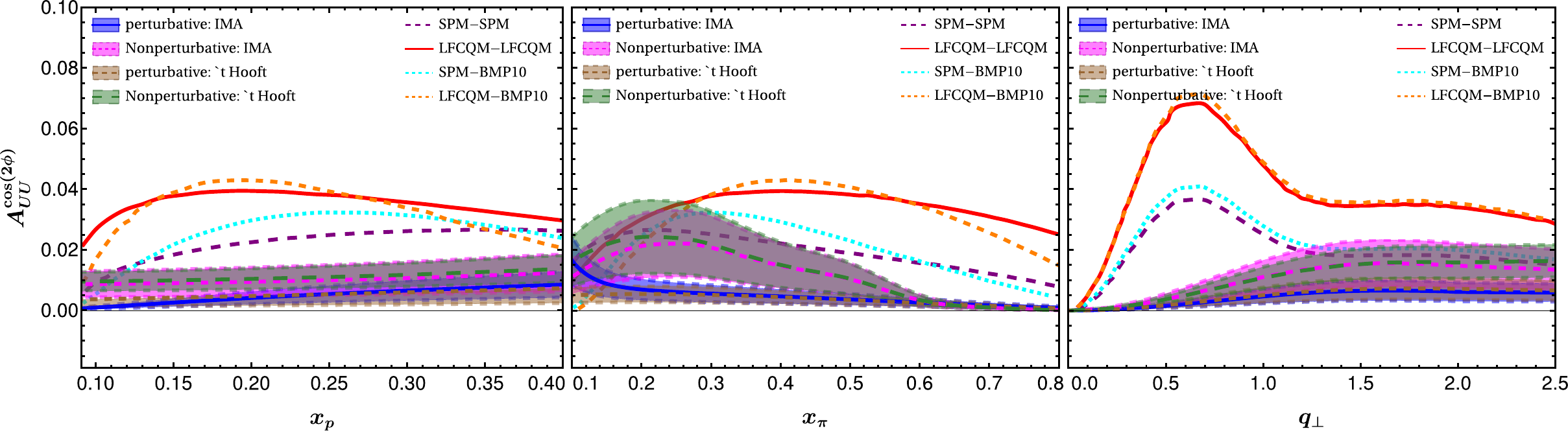}
	\caption{$\cos(2\phi)$ azimuthal asymmetry in pion-proton induced DY process. The panels from left to right show the variation of the azimuthal asymmetry with $x_{p}$, $x_{\pi}$, and $q_{\perp}$, 
 respectively. Our estimations (blue and magenta bands {are for IMA while brown and light-green bands are for 't Hooft}) are compared with the results obtained from pure model and hybrid calculations reported in Ref.~\cite{Bastami:2020asv}. The purple-dashed and solid-red lines show the pure model calculations from SPM~\cite{Gamberg:2009uk} and LFCQM~\cite{Pasquini:2014ppa} models, while the orange and cyan lines represent the hybrid computations of LFCQM and SPM models with BMP10 parametrization~\cite{Barone:2009hw}, respectively.}
	\label{fig:cos(2phi)} 
\end{figure*}

\begin{figure*}[ht]
	\centering
	\includegraphics[width=\textwidth]{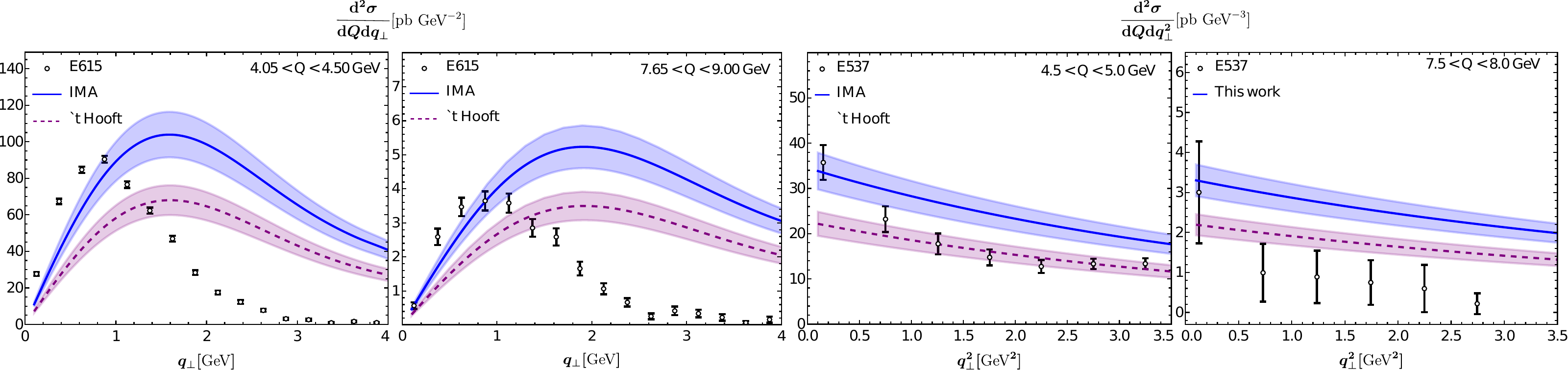}
	\caption{{Left panel: Comparison of pion-nucleus unpolarized DY differential cross-section as a function of $|\bfq|$ for IMA (solid blue line) and `t Hooft (purple dashed line) approaches with E615 dataset at center of mass energy $\sqrt{s}=21.8$ GeV for two different $Q$ bins. Right panel: Comparison of pion-nucleus unpolarized DY differential cross-section as a function of $\bfq^2$ for IMA and `t Hooft approaches with E537 dataset at center of mass energy $\sqrt{s}=15.3$ GeV for two different $Q$ bins.}}
	\label{fig:unpolcross} 
\end{figure*}
Figure~\ref{fig:sin(2phi+phis)} illustrates the $\sin(2\phi+\phi_{s})$ asymmetry, which is obtained by the convolution of pion Boer-Mulders function and the proton pretzelosity TMD, i.e., $A^{\sin(2\phi+\phi_{s})}_{UT}\propto h_{1,\pi}^{\perp\bar{q}}\otimes h_{1T,p}^{\perp q}$. 
We compare our model calculations with the available data~\cite{COMPASS:2017jbv,COMPASS:2023vqt}. This asymmetry has been studied theoretically in SPM and LFCQM~\cite{Gamberg:2009uk,Pasquini:2014ppa} and also studied with LP15 extracted proton pretzelosity distribution~\cite{Lefky:2014eia} along with SPM and LFCQM. This asymmetry is notably smaller compared to others, primarily because the pretzelosity TMD has a much smaller magnitude than other TMDs. Additionally, this asymmetry is proportional to $q_{\perp}^{3}$ for $q_{\perp}\ll 1$ GeV, while the other asymmetries are proportional to $q_{\perp}$. Consequently, this asymmetry is the smallest among the leading-twist asymmetries in pion-nucleon DY, typically amounting to 1\% or less numerically. To facilitate comparison with other model results, we provide insets indicating that the sign and magnitude of our asymmetry align with the SPM and LFCQM~\cite{Bastami:2020asv}, while the sign is opposite for the LP15 fit of pretzelosity~\cite{Lefky:2014eia}. The sign difference of pretzelosity distribution between model calculations and the first extraction from preliminary experimental data, suggest that the pion-induced DY process with a polarized proton opens a way to access the information on the pretzelosity distribution. Quantifying the pretzelosity TMD in DY and SIDIS presents inherent challenges. Nevertheless, the forthcoming high luminosity SIDIS experiments at JLab 12 GeV~\citep{Dudek:2012vr} and the prospective Electron-Ion Colliders~\citep{AbdulKhalek:2021gbh,Anderle:2021wcy} provide a promising avenue for the feasible measurement of pretzelosity.

In Figure~\ref{fig:sin(2phi)}, we present the $\sin(2\phi)$ longitudinal single-spin asymmetry. This asymmetry is obtained from the convolution of T-odd pion Boer-Mulders function and the proton Kotzinian-Mulders function, i.e., $A_{\textrm UL}^{\sin(2\phi)}\propto -h_{1,\pi}^{\perp\bar{q}}\otimes h_{1L,p}^{\perp q}$. We observe a positive distribution in our predicted asymmetry, indicating that the up quark Kotzinian-Mulders function is negative. This inference follows our earlier investigation~\cite{Gurjar:2023uho}, where we established the positivity of the pion Boer-Mulders function for the up quark. We find that our results are compatible with the predictions yielding from both the pure-model (LFCQM-LFCQM and SPM-SPM) and the hybrid calculations (LFCQM-WW-JAM20 and SPM-WW-JAM20)~\cite{Bastami:2020asv}. This asymmetry remains unmeasurable in COMPASS as it necessitates longitudinal polarization of the proton target for observation. Nonetheless, there is potential for future exploration of this phenomenon through DY experiments employing doubly polarized protons or deuterons, potentially within the framework of the NICA experiment~\cite{Savin:2014nth}.

In Figure~\ref{fig:cos(2phi)}, we show our predictions for the Boer-Mulders or  $\cos(2\phi)$ asymmetry in $\pi^{-}p$ unpolarized Drell-Yan process. This asymmetry is computed from the convolution of pion Boer-Mulders function and the proton Boer-Mulders function, i.e., $A_{UU}^{\cos(2\phi)}\propto h_{1,\pi}^{\perp \bar{q}}\otimes h_{1,p}^{\perp q}$. There is no experimental data from COMPASS for this asymmetry. However, a proposed strategy for analyzing this asymmetry has been outlined in ref.~\cite{COMPASS:2010shj}. We compare our predictions with pure models (SPM-SPM and LFCQM-LFCQM) as well as with hybrid calculations. In the hybrid calculations, the proton Boer-Mulders function is sourced from the BMP10 extraction~\cite{Barone:2009hw}, while the pion Boer-Mulders function is derived from both LFCQM~\cite{Pasquini:2014ppa} and SPM~\cite{Gamberg:2009uk}. We note that the magnitude of our predicted asymmetry is slightly smaller when compared to that of other theoretical predictions. Note that this asymmetry has also been investigated in ref.~\cite{Wang:2018naw}, where pion distributions are based on a light-cone model, while the proton distribution is adopted from the BMP10 extraction. Analyzing the results presented in Figure~\ref{fig:cos(2phi)} allows for the determination of the sign of the proton Boer-Mulders function. The positive distribution of the $\cos(2\phi)$ asymmetry across COMPASS kinematics leads to the inference that the proton Boer-Mulders function is positive in the DY process. This inference is supported by our prior discussion on the positivity of the pion Boer-Mulders function, from $A_{UT}^{\sin(2\phi-\phi_{s})}$ asymmetry~\citep{Gurjar:2023uho}. Notably, this observation aligns with the sign change observed in the process-dependent T-odd distribution function, consistent with findings in SIDIS analyses~\cite{Maji:2017wwd,Gurjar:2022rcl,Lyubovitskij:2022vcl}.
In Figures~\ref{fig:sin(phi)} - \ref{fig:cos(2phi)}, we illustrate the distinctions between the asymmetries produced through the utilization of perturbative and nonperturbative gluon rescattering kernels for the T-odd pion and proton TMDs. Our findings reveal that both the perturbatively and nonperturbatively generated asymmetries align reasonably well with the available experimental data. Furthermore, it is noteworthy that the asymmetries generated nonperturbatively exhibit a slightly larger magnitude compared to those generated perturbatively. We also note that both the holographic light-front wavefunctions with the IMA and the 't Hooft longitudinal model yield comparable spin asymmetries.
\section{Pion-nucleus induced DY  cross section}
{The leading order pion-nucleus Drell–Yan unpolarized differential cross-section can be expressed in terms of unpolarized structure function $F^{1}_{UU}$ as~\cite{Cerutti:2022lmb,Bacchetta:2022awv},
\begin{align}
    &\frac{{\rm d}\sigma^{DY}}{{\rm d}|\bfq|{\rm d}x_{F}{\rm d}Q}=\frac{16\pi^{2}\alpha^{2}_{\text{em}}|\bfq|}{9Qs\sqrt{x_{F}^{2}+4\frac{
    Q^{2}}{s}}}\sum_{q}e_{q}^{2}\int\frac{{\rm d}|\bfb||\bfb|}{2\pi}\nonumber\\
    &\times J_{0}(|\bfb||\bfq|) \hat{f}^{q}_{1,\pi}(x_{\pi},\bfb^{2};Q)\hat{f}^{\bar{q}}_{1,N}(x_{N},\bfb^{2};Q),
\end{align}
where $\alpha_{\text{em}}$ is the electromagnetic coupling, $x_{F}$ is the Feynman variable, $s=(P_{\pi}+P_{N})^{2}$ denotes the center-of-mass energy squared of pion-nuclei system and $Q=\sqrt{q^{2}}$ is the large invariant mass and $x_{\pi,N}=\pm\frac{x_{F}}{2}+\sqrt{\frac{x_{F}^{2}}{4}+\frac{Q^{2}}{s}}$ are the momentum fractions carried by the partons inside the pion and nucleon bound inside the nuclei. The impact parameter ($b_\perp$) dependent pion unpolarized TMD is $\hat{f}_{1,\pi}^{q}$ and $\hat{f}_{1,N}^{q}$ is the unpolarized impact parameter dependent nucleus TMD. The nuclei TMD distribution can be obtained by using the given expression~\citep{Hirai:2001np},
\begin{align}\label{nuclear_TMD}
    f_{1,N}^{u}(x_{N},\bfb^{2};Q)=&\frac{Z}{A}f_{1,p}^{u}(x_{p},\bfb^{2};Q)\nonumber\\
    &+\frac{A-Z}{A}f_{1,p}^{d}(x_{p},\bfb^{2};Q),
\end{align}
where $Z$ is the proton number and $A$ denotes the mass number of the nucleus. We neglect the European Muon Collaboration effect~\cite{EuropeanMuon:1983wih} and treat the target nucleus as a collection of independent nucleons.}

In Fig.~\ref{fig:unpolcross}, we present the cross sections for the pion-tungsten-induced DY process. For the tungsten nucleus, we set $Z=74$ and $A=184$ in Eq.~\eqref{nuclear_TMD}. Our model results are compared with data from two DY experiments at Fermilab, E615~\citep{Egido:1992ph} and E537~\citep{Anassontzis:1987hk}, at center of mass energies $\sqrt{s}=21.8$ GeV and $\sqrt{s}=15.3$ GeV for selected $Q$ bins, respectively. The solid blue line corresponds to the IMA approach, while the dashed purple line represents the 't Hooft approach. In the first two panels, the differential Drell-Yan cross-section is shown as a function of the transverse momentum $|\bfq|$ of the intermediate virtual photon, and compared with the E615 data. The last two panels display the differential Drell-Yan cross-section as a function of $|\bfq|^{2}$, with a comparison to the E537 data. We observe that, in magnitude, our model predictions align with the available experimental data, showing a broader distribution that peaks at larger values of $|\bfq|$. This discrepancy in width is likely due to our model including only the valence quark Fock sector contributions for the pion and proton TMDs, whereas the experimental data includes contributions from all quark flavors. Additionally, we note that the 't Hooft results are more consistent with the E537 dataset compared to the IMA results.
%
%
\section{Conclusion}\label{sec:conclusion}
%
We presented the azimuthal spin asymmetries, specifically $\sin(\phi_{s})$, $\sin(2\phi+\phi_{s})$, $\sin(2\phi)$, and $\cos(2\phi)$ asymmetries, in the single transversely polarized $\pi^-p$ Drell-Yan process within the TMD factorization formalism, focusing on the kinematics relevant to the COMPASS experiment.  These asymmetries arise from the convolution of the TMDs of the pion beam and the proton target. 
To describe the pion, we applied holographic light-front QCD models that differ in their treatment of longitudinal modes in the meson wave functions: (i) the IMA model, which lacks dynamical longitudinal modes, and (ii) a model incorporating longitudinal dynamics through the 't Hooft equation. Both approaches provide an excellent simultaneous description of a wide range of pion observables. Additionally, we employed the widely used quark-diquark model constructed by the soft-wall AdS/QCD for the proton. Gluon rescattering plays a critical role in obtaining a non-zero pion's and proton's T-odd TMDs. We explored the utilization of a nonperturbative SU$(3)$ gluon rescattering kernel, surpassing the conventional approximation of perturbative U$(1)$ gluons. 

After implementing the TMD evolution effect, we presented pure-model computations of the azimuthal asymmetries. 
Our analysis revealed that the asymmetries at COMPASS can be qualitatively depicted (in terms of sign and magnitude) by the current analysis of the pion's TMDs within the holographic light-front QCD framework and the proton's TMDs within a light-front quark-diquark model motivated by the soft-wall AdS/QCD. With respect to interpreting the initial data from the pion-induced Drell-Yan process with polarized protons, we discerned a robust understanding. Our analysis led to the conclusion of positive T-odd TMDs for the up quark in the proton within the context of the Drell-Yan process, which is opposite to the sign observed in SIDIS analyses~\cite{Kang:2009bp,Brodsky:2002cx,Brodsky:2002rv}. Based on the $\sin(2\phi+\phi_{s})$ asymmetry COMPASS data, no discernible trend is evident regarding the sign of the up quark's pretzelosity TMD. Nevertheless, we deduced a positive sign for this TMD, consistent with the predictions of the LFCQM and the SPM models, but contrasting with the LP15 fit of pretzelosity.
Due to the relatively large statistical uncertainties in COMPASS data, clear trends are not discernible for any of the azimuthal spin asymmetries. Upcoming, more precise data from COMPASS and other experimental facilities will enable us to further solidify the picture. We observed that the holographic light-front wavefunctions with both the IMA and the 't Hooft longitudinal mode produce similar spin asymmetries. We also examined the cross section for the pion-nucleus induced Drell-Yan process using the derived pion TMDs, along with the TMDs of the target nucleus, and found that the hLFQCD model incorporating the 't Hooft longitudinal mode offers a better fit to the experimental data than the hLFQCD model with the IMA. Our investigation has contributed to providing quantitative tests of the application of the holographic light-front QCD models to the description of the pion and the proton.
%
\section*{Acknowledgement}
%
The authors acknowledge fruitful discussions with B. Pasquini and D. Chakrabarti. BG also acknowledge Indian Institute of Technology Kanpur for providing the Fellowship for Academic and Research
Excellence (FARE). The work of CM is supported by new faculty start up funding by the Institute of Modern Physics, Chinese Academy of Sciences, Grant No. E129952YR0.  CM also thanks the Chinese Academy of Sciences Presidents International Fellowship Initiative for the support via Grants No. 2021PM0023. 
\bibliography{References.bib}
\end{document}